\documentclass{llncs}

\usepackage{hyperref}
\usepackage{tikz}
\usetikzlibrary{decorations.markings}
\usetikzlibrary{arrows}
\usepackage{amssymb}
\usepackage{amsfonts,amsmath,latexsym,stmaryrd}
\usepackage{listings}
\usepackage{cite}

\newcommand{\C}[1]{\mbox{\lstinline`#1`}}

\definecolor{dkblue}{rgb}{0,0.1,0.5} 
\definecolor{lightblue}{rgb}{0,0.5,0.5} 
\definecolor{dkgreen}{rgb}{0,0.4,0} 
\definecolor{dk2green}{rgb}{0.4,0,0} 
\definecolor{dkviolet}{rgb}{0.6,0,0.8}
\definecolor{shadethmcolor}{rgb}{0.9, 0.9,1}

\usepackage{verbatim}
\usepackage{rotating}
\usepackage{pdflscape}
\usepackage{booktabs}
\usepackage{theorem}
\usetikzlibrary{shapes,snakes}
\theorembodyfont{\rmfamily}

\newtheorem{PS}{Proof Strategy}

\begin{document}

\title{Proof-Pattern Search in Coq/SSReflect\thanks{The work was supported by EPSRC grants EP/J014222/1 and EP/K031864/1.}}

\author{J\'onathan Heras \and Ekaterina Komendantskaya}
\authorrunning{J. Heras and E. Komendantskaya}

\institute{School of Computing, University of Dundee, UK\\
\email{\{jonathanheras,katya\}@computing.dundee.ac.uk}}

\maketitle
 
\begin{abstract}

ML4PG is an extension of the Proof General interface, allowing
the user to invoke machine-learning algorithms and find proof similarities
in Coq/SSReect libraries. In this paper, we present three new improvements to
ML4PG.
First, a new method of \lq\lq{}recurrent clustering" is introduced to collect statistical features from Coq terms.
Now the user can receive suggestions about similar definitions,
types and lemma statements, in addition to proof strategies. Second, Coq
proofs are split into patches to capture proof strategies that could arise
at different stages of a proof. Finally, we improve ML4PG's output introducing
an automaton-shape representation for proof patterns. 
\end{abstract}

\section{Introduction}\label{sec:introduction}

Development of Interactive Theorem Provers (ITPs) has led to the creation of big libraries and varied infrastructures for formal
mathematical proofs. 
These frameworks usually involve thousands of definitions and theorems (for instance, there are approximately
4200 definitions and 15000 theorems in the formalisation of the Feit-Thompson theorem~\cite{FCT}). 
Parts of  those libraries can often be re-applied in new domains; however, 
it is a challenge for expert and non-expert users alike to trace them and find re-usable concepts and proof ideas.

Coq/SSReflect already provides comprehensive search mechanisms to search the corpus of results available in different libraries.
There are several search commands in Coq:  \lstinline?Search?~, \lstinline?SearchAbout?, \lstinline?SearchPattern?
and \lstinline?SearchRewrite?~\cite{Coq}. In addition, SSReflect implements its own version of the \lstinline?Search?
command \cite{SSReflect} -- SSReflect's \lstinline?Search? gathers the functionality of the 4 Coq's search commands.
The Whelp platform~\cite{AspertiGCTZ04} is a web search engine for mathematical knowledge formalised in Coq, which features 3 functionalities:
\lstinline?Match? (similar to Coq's \lstinline?Search? command), \lstinline?Hint? (that finds all the theorems which can
be applied to derive the current goal) and \lstinline?Elim? (that retrieves all the eliminators of a given type).

The existing search mechanisms can be used in two different scenarios. If the user knows how to continue the proof, but
he does not remember (or know) the concrete name of the desired auxiliary lemma, it suffices to provide a search pattern to Coq searching engines, e.g.
using commands of the form
``\lstinline?Search "distr" in bigop?'' or   
``\lstinline?Search _ (_ * (\big[_/_]_(_ <- _| _) _))?'',
 where \texttt{bigop} is a library, \texttt{"distr"} is a pattern in a lemma name, and  \lstinline? _ (_ * (\big[_/_]_(_ <- _| _) _))?
is a pattern for search.
 
In the second scenario, the user needs help in the middle of the proof. Searching mechanisms can be also useful in this 
case; for instance, using the \lstinline?Search? command the user can search all the lemmas associated with a concrete term
of the current goal. The \lstinline?Hint? mechanism of Whelp can find all the theorems which can be applied to derive the current goal.



The above search mechanisms are goal directed and deterministic. That is, the user
searches  the chosen  libraries for lemmas related to a concrete type, term or pattern. 
If the patterns defined by the user are present in the given library, then the user is guaranteed to see the relevant lemmas on the screen.

The situation is more complicated if the user does not know the right pattern to search for. Imagine, for example, being in the middle of constructing a proof, 
and wishing to get some higher-level hint on how to proceed, wishing there was an ITP expert near, who would suggest a further proof strategy based on his previous experience.
The ML4PG (``Machine-Learning for Proof-General'') tool~\cite{KHG13,CICM13,HK14} was created to emulate such intelligent help automatically, using statistical machine-learning
algorithms to use the information arising from the ``previous experience'' with other proofs.

The main idea was to use statistical clustering to detect common proof-patterns. A proof pattern was defined as a correlation between the tactics  and the types/shapes of 
subgoals resulting from the tactic applications, 
within a few proof steps.
The resulting tool could indeed find some interesting -- unexpected and yet relevant -- proof patterns,
across different notation, types, and libraries. Our experiments span several subjects -- basic mathematical infrastructures, Computer Algebra, Game Theory, and
certification of Java Virtual Machine; the results are best summarised in~\cite{CICM13,HK14}.

In comparison to the more traditional \emph{search engines}, ML4PG is a goal-independent (unsupervised) tool, i.e. the user does not have to know the required pattern in advance.
But, as many 
statistical machine-learning applications, it is also non-deterministic. That is, the tool failing to suggest a proof pattern does not mean there is no ``interesting'' pattern to be found.
Actually, the notion of a proof pattern being ``interesting'' is left to the user's judgement, as well.

That initial approach had one inherent limitation: the essence of a Coq/SSReflect proof is not fully expressible by a sequence of applied tactics. 
The definitions, types, and shapes of auxiliary lemmas used in a proof can be much more sophisticated and conceptual than a proof script calling them.
Therefore, although ML4PG could find interesting, and often useful, sequences of tactics; it could not go further to recognise e.g. similar definitions.
 
In this paper, we present the most recent extensions to ML4PG,  involving all kinds of Coq terms -- type declarations, definitions and lemma statements -- into
the process 
of pattern search. This required additional algorithms of feature extraction that reflect the mutual dependency of various proof objects in Coq\rq{}s dependently-typed setting; see Sections~\ref{sec:lemmaclustering} and~\ref{sec:reclemmaclustering}. 
This major step in ML4PG development prompted other improvements. The initial ML4PG was considering features arising from first 5 proof steps in a proof, whereas now we treat every proof as a collection of
proof patches, each potentially
representing an interesting proof strategy. Moreover, if say 15th-20th step in one proof resembles a 115th-120th step in another, the tool is now able to detected such patterns deep down.  
The feature extraction algorithms for proof features have been further refined to include the data collected from Coq terms, and now the 
whole syntax of the chosen proof libraries is subject
to \emph{recurrent clustering} -- a novel technique for ML4PG. 
All these extensions to proof-feature extraction are explained in Section~\ref{sec:recurrent}. 

We also improve conceptualisation of ML4PG's output. It used to show, in response to the user's call, a set of similar proofs,
with no hints of why these proofs are deemed similar. We now introduce an automaton-shape 
representation of each detected proof-patch, showing the proof-features that correlate, see Section~\ref{sec:conceptualisation}. This partially addresses the drawback of the subjective approach to 
the pattern's ``interestingness'' -- now 
the tool clearly declares correlation of which proof features defined the suggested proof pattern. 
Finally, in Section~\ref{sec:conclusions}, we conclude the paper.
Examples we use throughout the paper come from several SSReflect libraries: the basic infrastructure of SSReflect~\cite{SSReflect}, a matrix library~\cite{GarillotEtAl09}
and a formalisation of persistent homology~\cite{HCMS12}. ML4PG is a part of standard Proof General distribution, the novel features we present here are available at~\cite{HK12}. 

\section{Feature Extraction for Coq terms}\label{sec:lemmaclustering}

ML4PG uses (unsupervised) \emph{clustering algorithms}~\cite{Bishop} to find patterns in Coq syntax and proofs.
Clustering algorithms 
divide data into $n$ groups of similar objects (called \emph{clusters}), where the value of $n$
is a parameter provided by the user. In ML4PG, the value of $n$ is automatically computed depending on the number 
of objects to cluster, and using the formula provided in~\cite{KHG13,lpar13}. 
A detailed exposition of machine-learning algorithms involved in ML4PG can be found in \cite{KHG13}. In this paper, we focus on improving ML4PG's feature extraction.

\emph{Feature extraction}~\cite{Bishop} is a research area developing methods for 
discovery of statistically significant features in data. 
We adopt the following standard terminology. 
We assume there is a training data set, containing some samples (or objects). 
\emph{Features} are given by a set of statistical parameters chosen to represent all objects in the given data set.
If $n$ features are chosen, one says that object classification is 
conducted in an $n$-dimensional space. 
For this reason, most pattern-recognition tools will require that 
the number of selected features is limited and fixed (sparse methods, like the ones applied in
e.g.~\cite{lpar-urban,mash,UrbanSPV08}, are the exception to the rule). 
\emph{Feature values} are rational numbers used to instantiate the features for every given object.  
If an object is characterised by $n$ feature values, these $n$ values together form a \emph{feature vector} for this object. 
A function that assigns, to every object of the data set, a feature vector is called a \emph{feature extraction function}.
Normally, feature extraction is a data pre-processing stage, separate from the actual pattern-recognition. 

Feature extraction from terms or \emph{term trees} is common to most feature extraction algorithms implemented
in theorem provers~\cite{lpar13,lpar-urban,mash,UrbanSPV08}. In~\cite{lpar13}, we introduced a feature extraction mechanism
for ACL2 first-order terms. Here, that ACL2 method is substantially re-defined to capture the higher-order dependently-typed language of Coq.

The underlying formal language of Coq is known as the \emph{Predicative Calculus of (Co)Inductive Constructions} (pCIC)~\cite{Coq,CoquandH88,CoPa89}. 
The terms of pCIC are built from
the following rules~\cite{Coq,CoquandH88,CoPa89}:

\begin{definition}[pCIC term]\label{def:coqterms}

$-$ The sorts \lstinline?Set?, \lstinline?Prop?, \lstinline?Type(i)? (\lstinline?i?$\in \mathbb{N}$) are terms.

$-$ The global names of the environment are terms. 

$-$ Variables are terms.
 
$-$ If \lstinline?x? is a variable and \lstinline?T, U? are terms; then, \lstinline?forall x:T,U? 
 is a term. If \lstinline?x? does not occur in \lstinline?U?; then, \lstinline?forall x:T,U? will be written as \lstinline?T -> U?. A term of the 
 form \lstinline?forall x1:T1, forall x2:T2, ..., forall xn:Tn, U? will be written as \lstinline?forall (x1:T1) (x2:T2) ...(xn:Tn), U?. 
 
$-$ If \lstinline?x? is a variable and \lstinline?T, U? are terms; then, \lstinline?fun x:T => U? 
 is a term. A term of the form \lstinline?fun x1:T1 => fun x2:T2 => ... => fun xn:Tn => U? will be written as \lstinline?fun (x1:T1) (x2:T2) ...(xn:Tn) => U?. 

 $-$ If \lstinline?T? and \lstinline?U? are terms; then, \lstinline?(T U)? is a term -- we use an uncurried notation (\lstinline?(T U1 U2 ... Un)?) 
 for nested applications (\lstinline?(((T U1) U2) ... Un)?).
 
 $-$ If \lstinline?x? is a variable, and \lstinline?T, U? are terms; then, (\lstinline?let x:=T in U?) is a term.

\end{definition}

The syntax of Coq terms~\cite{Coq} includes some terms that do not appear in Definition~\ref{def:coqterms}; e.g. given 
 a variable \lstinline?x?, and  terms \lstinline?T? and \lstinline?U?, \lstinline?fix name (x:T) := U? is a Coq term used to declare a recursive definition.
The notion of a term in Coq covers a very general syntactic
category in the Gallina specification language~\cite{Coq} and corresponds to the intuitive notion of well-formed expression. 
However, for the purpose of concise exposition, 
we will restrict our notion of a term to Definition~\ref{def:coqterms} in this paper,
giving the full treatment of the whole Coq syntax in the actual ML4PG implementation.

\begin{definition}[ML4PG term tree]
Given a Coq term \lstinline?C?, we define its associated term tree as follows:

$-$ If \lstinline?C? is one of the sorts \lstinline?Set?, \lstinline?Prop? or \lstinline?Type(i)?; then, the term tree of 
 \lstinline?C? consists of one single node, labelled respectively by \lstinline?Set:Type(0)?, \lstinline?Prop:Type(0)? or \lstinline?Type(i):Type(i+1)?.
 
$-$ If \lstinline?C? is a name or a variable; then, 
 the term tree of \lstinline?C? consists of one single node, labelled by the name or the variable itself together with its type.
 
$-$ If \lstinline?C? is a term of the form \lstinline?forall (x1:T1) (x2:T2) ...(xn:Tn), U? (analogously for \lstinline?fun (x1:T1) (x2:T2) ...(xn:Tn) => U?); 
then, the term tree of \lstinline?C? is the tree with the root node labelled by \lstinline?forall? (respectively \lstinline?fun?) 
and its immediate subtrees given by the trees representing \lstinline?x1:T1?, \lstinline?x2:T2?, \lstinline?xn:Tn? and \lstinline?U?.

$-$ If \lstinline?C? is a term of the form \lstinline?let x:=T in U?; then, the term tree of \lstinline?C? 
 is the tree with the root node labelled by \lstinline?let?, having three subtrees given by the trees corresponding to \lstinline?x?, \lstinline?T? and \lstinline?U?.

$-$ If \lstinline?C? is a term of the form  \lstinline?T -> U?; then,  the term tree of \lstinline?C?  is represented by the tree with the root node labelled by 
\lstinline?->?, and its immediate subtrees given by the trees representing  \lstinline?T? and \lstinline?U?.

$-$ If \lstinline?C? is a term of the form  \lstinline?(T U1 ... Un)?; then, we have two cases.
If \lstinline?T? is a name, the term tree of \lstinline?C?  is represented by 
the tree with the root node labelled by \lstinline?T? together with its type, and its immediate subtrees given by the trees
representing \lstinline?U1?,\ldots, \lstinline?Un?. If \lstinline?T? is not a  name, the term tree of \lstinline?C?
is the tree with the root node labelled by \lstinline?@?, and its immediate subtrees given by the trees 
representing \lstinline?T?, \lstinline?U1,...,Un?.

\end{definition}

Note that ML4PG term trees consist of two kinds of nodes: \emph{Gallina} and \emph{term-type} nodes. The Gallina nodes are 
labelled by Gallina keywords and special tokens such as \lstinline?forall?, \lstinline?fun?, \lstinline?let? or \lstinline?->? (from now on, we will call them Gallina tokens) 
and, the term-type nodes are labelled by expressions of the form \lstinline?t1:t2? where \lstinline?t1? is a sort, a variable or 
a name, and  \lstinline?t2? is the type of \lstinline?t1?.

\begin{figure}[t]
\centering
\begin{tikzpicture}[level 1/.style={sibling distance=40mm},
 level 2/.style={sibling distance=70mm},
 level 3/.style={sibling distance=30mm},
 level 4/.style={sibling distance=30mm},scale=.8,font=\footnotesize]
   
   \node (root) {\lstinline?forall?} [level distance=10mm]
             child { node {\lstinline?n : nat?}}
             child { node {\lstinline?H : even n?}}
             child { node {\lstinline?odd : nat -> Prop?} 
                      child { node {\lstinline?+ : nat -> nat -> nat?}
                           child { node {\lstinline?n : nat?}}
                           child { node {\lstinline?1 : nat?}}
                           }
                   }
            
    ;
 \end{tikzpicture}
\caption{\scriptsize{\emph{ML4PG term tree for \texttt{forall (n : nat) (H : even n), odd (n + 1).}}}}\label{fig:termtree}
\end{figure}  

\begin{example}\label{ex1}
 Given the term \lstinline?forall (n : nat) (H : even n), odd (n + 1)?, its ML4PG term tree is depicted in Figure~\ref{fig:termtree}.
 \end{example}

We represent ML4PG term trees as feature matrices, further to be flattened into feature vectors for clustering. 
A variety of methods exists to represent trees as matrices, for instance using
adjacency or incidence matrices. The adjacency matrix and the various previous methods of  feature
extraction (e.g.~\cite{lpar-urban,mash,UrbanSPV08}) share the following common properties: different 
library symbols are represented by distinct features, and the 
feature values are binary. 
For
big libraries and growing syntax, feature vectors grow very large (up to $10^6$ in some experiments)
and at the same time very sparse, which implies the use of sparse machine-learning 
in~\cite{lpar-urban,mash,UrbanSPV08}.


We develop a new compact method that tracks a large (potentially unlimited) number of Coq terms by a finite number of features, and unlimited number of feature values.
In our method, the features are given by two properties common to all possible term trees: the term tree depth and the level index of nodes.
The most important information about the term is then encoded by improving precision of feature values 
using rational-valued feature-extraction functions. Taking just $300$ features, the new feature-extraction method recursively adjusts the feature values, 
adapting to the growing language syntax. 
The resulting feature vectors have density ratio of 60\%.   

Given a Coq expression, we can differentiate its term and type components; the feature values capture information from
these components and also the structure of the tree. In particular, each tree node is encoded by distinct feature values
given by a triple of rational numbers to represent the term component, the type component, and the level index of the parent node in the term tree,
cf. Table~\ref{ml4pgtermtable}. Our feature extraction method is formalised in the following definitions.

\begin{table}[t]
\centering
{\scriptsize 
\begin{tabular}{|c||c|c|c|}
\hline
 &  level index 0 & level index 1 & level index 2 \\
 \hline
td0 & ($[\texttt{forall}]_{Gallina}$,-1,-1) & (0,0,0) & (0,0,0)\\
\hline
td1 & ($[\texttt{n}]_{term}$,$[\texttt{nat}]_{type}$,0) & ($[\texttt{H}]_{term}$,$[\texttt{even n}]_{type}$,0)& ($[\texttt{odd}]_{term}$,$[\texttt{nat-> Prop}]_{type}$,0) \\
\hline
td2 & ($[\texttt{+}]_{term}$,$[\texttt{nat -> nat -> nat}]_{type}$,2) & (0,0,0)& (0,0,0)\\
\hline
td3 & ($[\texttt{n}]_{term}$,$[\texttt{nat}]_{type}$,0) & ($[\texttt{1}]_{term}$,$[\texttt{nat}]_{type}$,0)  & (0,0,0)\\
\hline
\end{tabular}}
\caption{\scriptsize{\emph{ML4PG term tree matrix for \texttt{forall (n : nat) (H : even n), odd (n+1)}.}}}\label{ml4pgtermtable}
\end{table}

\begin{definition}[Term tree depth level and level index]\label{def:termtreelevel}
Given  a term tree $T$, the \emph{depth} of the node $t$ in $T$, denoted by \emph{depth(t)}, is defined as follows:

$-$ $depth(t) = 0$, if $t$ is a root node;

$-$ $depth(t) = n+1$, where $n$ is the depth of the parent node of $t$.

The \emph{$n$th level} of $T$ is the ordered sequence of nodes of depth $n$ -- using the classical representation for trees, the order of the sequence is 
given by visiting the nodes of depth $n$ from left to right. The \emph{level index} of a node with depth $n$ is the position of the node in the $n$th level of $T$.
We denote $T(i,j)$ to the node of $T$ with depth $i$ and index level $j$.
\end{definition}
	
	
%
%
%

We use the notation $M[\mathbb{Q}]_{n\times m}$ to denote the set of matrices of size  $n\times m$ with rational coefficients.

\begin{definition}[ML4PG term tree feature extraction]\label{df:matrix}
Given a term \lstinline?t?, its corresponding term tree $T_{\texttt{t}}$, and three injective functions 
$[.]_{term}: Coq~terms \rightarrow \mathbb{Q}^+$, $[.]_{type}: Coq~terms \rightarrow \mathbb{Q}^+$
and $[.]_{Gallina}: Gallina~tokens \rightarrow \mathbb{Q}^-$; then, the feature extraction function 
$[.]_M=<[.]_{term}, [.]_{type},[.]_{Gallina}> : Coq~terms \rightarrow M[\mathbb{Q}]_{10\times 10}$
builds the  \emph{term tree matrix of \texttt{t}}, $[\texttt{t}]_M$,
where the $(i,j)$-th entry of $[\texttt{t}]_M$ captures information from the node $T_{\texttt{t}}(i,j)$ as follows:

$-$ if $T_{\texttt{t}}(i,j)$ is a Gallina node $g$; then, the $(i,j)$th entry of $[\texttt{t}]_M$ is a triple $([g]_{Gallina},-1,p)$ 
where $p$ is the level index of the parent of $g$. 

$-$ if $T_{\texttt{t}}(i,j)$ is a term-type node \lstinline?t1:t2?; then, the $(i,j)$th entry of $[\texttt{t}]_M$ is a triple $([t1]_{term},[t2]_{type},p)$
where $p$ is the level index of the parent of the node.

\end{definition}

In the above definition, we fix  the maximum depth and  maximum level index of a node to $10$; this makes the feature extraction mechanism uniform
across all Coq terms appearing in the libraries. We may lose some information if  pruning is needed, but the chosen size works well for most terms appearing in Coq libraries. 
If a term tree does not fit into $10 \times 10$ term tree dimensions, its $10 \times 10$
subtree is still considered by ML4PG.
The term tree matrix is flattened into a feature vector and each triple will be split 
into three components of the vector, giving a feature vector size of $300$, 
still smaller than in sparse approaches~\cite{lpar-urban,mash,UrbanSPV08}.

In Definition~\ref{df:matrix}, we deliberately specify the functions $[.]_{Gallina}, [.]_{term}$ and $[.]_{type}$ just by their signature. 
The function $[.]_{Gallina}$ is a predefined function.
The number of Gallina tokens (\lstinline?forall?, \lstinline?fun?, \lstinline?->? and so on) is fixed and cannot be 
expanded by the Coq user. Therefore, we know in advance all the Gallina tokens that can appear in a development, and we can
assign a concrete value to each of them. The function $[.]_{Gallina}: Gallina~tokens  \rightarrow \mathbb{Q}^-$ is an injective 
function carefully defined to assign close values to similar Gallina tokens 
and more distant numbers to unrelated tokens 
-- see Appendix~\ref{sec:gallinasyntax} for the exact encoding.

The functions $[.]_{term}$ and $[.]_{type}$ are dynamically re-defined for every 
library and every given proof stage, to adapt to the changing syntax.
In practice, there will be  
new $[.]_{term}$ and $[.]_{type}$ functions computed whenever ML4PG is called.
This
brings the element of the ``acquired knowledge/experience'' to the machine-learning cycle, as will be formalised in the next section.




\section{Recurrent Term Clustering}\label{sec:reclemmaclustering}

The previous section introduced a method of defining statistically significant features. 
It remains to define the functions
 $[.]_{term}$ and $[.]_{type}$ that will determine feature values.
These functions must be sensitive to the structure of terms, assigning close values 
to similar terms and more distant values to unrelated terms. 

A term \lstinline?t? is represented by $300$ feature values of $[\texttt{t}]_M$.
The values of $[.]_{M}$
for  variables  and pre-defined sorts in \lstinline?t? are fixed, but the values of user-defined terms (and types!) contained in \texttt{t}
have to be computed recursively, based
on the structures of their definitions, and 
using clustering to compute their feature vectors, and their representative values for $[\texttt{t}]_M$. 
It is the nature of functional languages to have terms depending on other terms, and feature extraction/clustering cycle is 
repeated recursively to reflect complex mutual term dependencies as feature values.  
 We call this method \emph{recurrent clustering}: the function $[.]_M$ automatically (and recurrently) adapts to the given libraries and the current proof stage. 
%
This differs from the standard machine-learning approach (and the old version of ML4PG), where the process of feature extraction is separated from running pattern-recognition algorithms. Here, one is a crucial part of another.

When Coq objects are divided into clusters, a unique integer number is assigned to each cluster. Clustering algorithms compute 
a \emph{proximity value} (ranging from $0$ to $1$) to every object in a cluster to indicate the certainty of the given example belonging 
to the cluster. The cluster numbers and the proximity values are used in the definitions of $[.]_{term}$ and $[.]_{type}$ below.

\begin{definition}\label{def:funterm}
Given a term  \lstinline?t? of a Coq library, the functions $[.]_{term}$ and $[.]_{type}$ are defined respectively for the term component \lstinline?t1?
and the type component \lstinline?t2? of every term-type node in the ML4PG term tree of \lstinline?t? as follows:

$-$ $[\texttt{t1}]_{term/type}=i$, if \lstinline?t1? is the $i$th distinct variable in \lstinline?t?. 

$-$ $[\texttt{t1}]_{term/type}=100+\sum_{j=1}^i\frac{1}{10\times 2^{j-1}}$, if \lstinline?t1? is the $i$th element of the set\\ $\{\texttt{Set},\texttt{Prop},\texttt{Type(0)},
\texttt{Type(1)}, \texttt{Type(2)},\ldots\}$.
 
$-$ $[\texttt{t1}]_{term}=200+2\times j + p$, where $j$ is a number of a cluster $C_j$ computed by the latest run of term clustering, 
such that $p$ is the proximity value of $\texttt{t1}$ in $C_j$. 


 
$-$ $[\texttt{t2}]_{type}=200+2\times j + p$, where $j$ is a number of a cluster $C_j$ computed by the latest run of type clustering (i.e. term clustering restricted to types),
such that $p$ is the proximity value of $\texttt{t2}$ in $C_j$.
%
\end{definition}

Note the recurrent nature of the functions $[.]_{term}$ and $[.]_{type}$ where numbering of components of \lstinline?t? depends on the term definitions
and types included in the library, assuming those values are computed by iterating the process back to the basic definitions.  In addition, the function $[.]_{term}$ internally uses the function $[.]_{type}$ in the recurrent clustering 
process and \emph{vice versa}. 

In the above definition, 
the variable encoding reflects the number and order of unique variables appearing in the term, note its similarity to the De Bruijn indexes.
In the formula for sorts, $\sum_{j=1}^i\frac{1}{10\times 2^{j-1}}$ reflects the close relation among sorts, and 
$100$ is used to  distinguish sorts from variables and names. 
Finally, the formula $200+2\times j + p$ assigns $[\texttt{t1}]$ (or $[\texttt{t2}]$) a value within $[200+2\times j,200+2\times j+1]$ depending on the
statistical proximity of \lstinline?t1? (or \lstinline?t2?) in cluster $j$. Thus, elements of the same cluster have closer values comparing to the values 
assigned to elements of other clusters, sorts, and variables. The formula is the same for the functions $[.]_{term}$ and $[.]_{type}$, but it is computed with different 
clusters and the values 
occur in different cells of the term tree matrices (cf. Definition~\ref{df:matrix}); thus, clustering algorithms distinguish terms and types on the level of features rather than feature values.

We can now state the main property of the ML4PG feature extraction.

\begin{proposition}
Let ${\cal T}$ be the set of Coq terms  whose trees have maximum depth $10$ and level index $10$. 
Then, the function $[.]_{M}$ restricted to ${\cal T}$ is a one-to-one function. 
\end{proposition}

Once the feature values of ML4PG term tree matrices have been computed, we can cluster these matrices and 
obtain groups of similar terms. In particular, ML4PG can be used to cluster definitions, types and lemma 
statements. 
We finish this section with some clusters discovered among the 457 definitions of the basic infrastructure of the SSReflect 
library.

\begin{example}
We include here 3 of the 91 clusters discovered by ML4PG automatically in the SSReflect library of 457 terms (across 12 standard files), within 5-10 seconds.
Note that this example of cluster-search is not goal-oriented, ML4PG discovers patterns without any user guidance, and offers the user to consider term similarities of which he may 
not be aware. 

\noindent - Cluster 1:
{\scriptsize \begin{lstlisting}
 Fixpoint eqn (m n : nat) :=
   match m, n with 
   | 0, 0 => true | m'.+1, n'.+1 => eqn m' n' 
   | _, _ => false end.
 Fixpoint eqseq (s1 s2 : seq T)  :=
   match s1, s2 with 
   | [::], [::] => true | x1 :: s1', x2 :: s2' => (x1 == x2) && eqseq s1' s2' 
   | _, _ => false end.         
\end{lstlisting}}
    
\noindent - Cluster 2:
{\scriptsize 
\begin{lstlisting}
 Fixpoint drop n s := match s, n with | _ :: s', n'.+1 => drop n' s' | _, _ => s end.
 Fixpoint take n s := match s, n with | x :: s', n'.+1 => x :: take n' s' | _, _ => [::] end.
\end{lstlisting}}

\noindent - Cluster 3:
{\scriptsize 
\begin{lstlisting}
 Definition flatten := foldr cat (Nil T).
 Definition sumn := foldr addn 0.
\end{lstlisting}} 
\end{example}

The first cluster contains the definitions of equality for natural numbers and lists -- showing that 
ML4PG can spot similarities across libraries. The second cluster discovers the relation between \lstinline?take? (takes the first $n$ elements of 
a list) and \lstinline?drop? (drops the first $n$ elements of a list). 
The last pattern is less trivial of the three, as it depends on 
other definitions, like \lstinline?foldr?, \lstinline?cat? (concatenation of lists) and \lstinline?addn? (sum
of natural numbers). 
Recurrent term clustering handles such dependencies well: it assigns close values to \lstinline?cat? and \lstinline?addn?,  since they have been discovered to belong to the same cluster. 
Note the precision of ML4PG clustering. Among $457$ terms it considered, $15$ used \lstinline?foldr?, however,  Cluster 3 contained only $2$ definitions, excluding e.g. 
\lstinline?Definition allpairs s t:=foldr (fun x => cat (map (f x) t)) [::] s? ; \lstinline?Definition divisors n:=foldr add_divisors [:: 1] (prime_decomp n)?  or \lstinline?Definition Poly:=foldr cons_poly 0.?

To summarise, there are three main properties that distinguish ML4PG pattern search from standard Coq search commands:
\begin{itemize}
	\item the user does not have to know and provide any search pattern;
	\item the discovered clusters do not have to follow a \lq\lq{}pattern\rq\rq{} in a strict sense (e.g. neither exact symbol names nor their order make a pattern), but ML4PG considers structures and background information found in the library; and,
	\item working with potentially huge sets of Coq objects, ML4PG makes its own intelligent discrimination of more significant and less significant patterns, 
	as example with \lstinline?foldr? has shown. This is opposed to the classic search for \lstinline?foldr? pattern that would give you a set of $15$ definitions.
\end{itemize}

ML4PG can work in a goal-directed mode, 
and discover only clusters of terms that are similar to the given term \lstinline?t?.
This can speed-up the proof development in two different ways. 
Clustering will provide definitions of terms similar to \lstinline?t?; 
hence, the proofs of the theorems 
involving those terms may follow similar patterns.
Clustering can also discover that a newly defined term \lstinline?t? was previously defined (perhaps in a different notation, as ML4PG works with structures across notations);
in that case, the user can use the existing library definition and all its background theory instead of defining it from scratch.

\section{Recurrent Proof Clustering}\label{sec:recurrent}

The method presented in the previous section can cluster similar statements of all Coq terms, including lemmas and theorems. 
However, this method does not capture the interactive nature of Coq proofs.
In this section, we involve proofs into the recurrent clustering of Coq libraries.

 In~\cite{KHG13}, 
we introduced a feature extraction method for Coq proofs capturing
the user's interaction through the applied tactics. That method 
traced 
low-level properties present in proof's subgoals, e.g. ``the
top symbol'' or ``the argument type''. Further, these features were taken
in relation to the statistics of user actions on every subgoal: how many and what
kind of tactics he applied, and what kind of arguments he provided to the
tactics. Finally, a few proof-steps were taken in relation to each other.
This method  had two drawbacks.\\ 
(1) It was  focused on the first
five proof-steps of a proof; therefore, some information was lost. We address this issue by implementing automatic split of each proof into proof-patches, thus allowing ML4PG to analyse a proof by the properties
of the patches that constitute the proof. \\
(2) The method assigned most feature values blindly, thus being insensitive to many important parameters, such as e.g. the structure of lemmas and hypotheses used as 
tactic arguments within a proof. 
Last section gave us the way of involving all Coq objects into recurrent feature re-evaluation.



\begin{definition}[Coq proof]
 Given a statement $S$ in Coq, a \emph{Coq proof} of $S$ is given by a sequence of triples $((\Gamma_i,G_i,T_i))_{0\leq i\leq n}$ where $\Gamma_i$ is a context,
 $G_i$ is a goal and $T_i$ is a sequence of tactics satisfying:

  - $G_0=S$, and $\forall i$, $\Gamma_i$ is the context of the goal $G_i$,
  
  - $\forall i$ with $0<i\leq n$, $\Gamma_i,G_i$ are respectively the context and goal obtained after applying $T_{i-1}$, and 
    the application of $T_n$ completes the proof.
\end{definition}

In this paper, we focus on the goals and tactics of Coq proofs; thus, we do not consider the contexts and denote the 
Coq proof $((\Gamma_i,G_i,T_i))_{0\leq i\leq n}$ by $((G_i,T_i))_{0\leq i\leq n}$. Involving contexts into proof-pattern search may be a subject for future work.

\begin{table}[t]
 	\centering
 	\tiny{
 		\begin{tabular}{|l|l|}
 		\hline
 	Goals and Subgoals & Applied Tactics \\
 		\hline
 		\hline
 	{\scriptsize $G_0) \forall~n,\sum\limits_{i=0}^{n} (g(i+1) - g(i)) = g(n+1) - g(0)$} & \\
 			& $T_0)$ {\scriptsize \lstinline?case : n => [|n _].?} \\
        $G_1) \sum\limits_{i=0}^{0} (g(i+1) - g(i)) = g(1) - g(0)$ & \\
        & $T_1)$  {\scriptsize\lstinline?by rewrite big_nat1.?}\\
        $G_2)  \sum\limits_{i=0}^{n+1} (g(i+1) - g(i)) = g(n+2) - g(0)$ &\\
        
        & $T_2)$ {\scriptsize\lstinline?rewrite sumrB big_nat_recr big_nat_recl ?}\\
          & ~~~~~~~~{\scriptsize\lstinline?     addrC addrC -subr_sub -!addrA addrA.?} \\
        $G_3) g(n+2) + \sum\limits_{i=0}^{n} g(i+1) -  \sum\limits_{i=0}^{n} g(i+1) - g(0) =$ &\\
        $  g(n+2) - g(0)$ &\\
        &$T_3)$  {\scriptsize\lstinline?move : eq_refl.?}\\
        $G_4) \sum\limits_{i=0}^{n} g(i+1) == \sum\limits_{i=0}^{n} g(i+1) \rightarrow  $ &\\
        $g(n+2) + \sum\limits_{i=0}^{n} g(i+1) -  \sum\limits_{i=0}^{n} g(i+1) - g(0) =$ &\\
        $  g(n+2) - g(0)$ &\\
        &$T_4)$  {\scriptsize\lstinline?rewrite -subr_eq0.?}\\
        $G_5) \sum\limits_{i=0}^{n} g(i+1) - \sum\limits_{i=0}^{n} g(i+1) == 0\rightarrow  $ &\\
        $g(n+2) + \sum\limits_{i=0}^{n} g(i+1) -  \sum\limits_{i=0}^{n} g(i+1) - g(0) =$ &\\
        $  g(n+2) - g(0)$ &\\
        &$T_5)$  {\scriptsize\lstinline?move/eqP => ->.?}\\
         $G_6)  g(n+2) + 0 - g(0) = g(n+2) - g(0)$ &\\
        &$T_6)$  {\scriptsize\lstinline?by rewrite sub0r.?}\\
 		$\Box$ & \\
 		& {\scriptsize\lstinline?Qed.?}\\
 		\hline
 		\end{tabular}
 		
 		}
 	\caption{\scriptsize{\emph{Proof for the lemma of Example~\ref{example0} in SSReflect.}}}
 	\label{tab:sumfirstn}
 \end{table}

\begin{example}\label{example0}
Table~\ref{tab:sumfirstn} shows the Coq proof of the following statement:
 $$\forall g:\mathbb{N} \rightarrow \mathbb{Z}\implies \sum_{0\leq i \leq n} (g(i+1) - g(i)) = g(n+1) - g(0)$$
\end{example}

One small proof may potentially 
resemble a fragment of a bigger proof; also, various small ``patches'' of different big proofs may resemble. 

\begin{definition}[Proof-patch]
 Given a \emph{Coq proof} $C=((G_i,T_i))_{0\leq i\leq n}$, a \emph{proof-patch} of $C$ is a subsequence of at most $5$ consecutive 
 pairs of $C$.
\end{definition}

From proof-patches, we can construct the feature matrices. 
We will shortly define the feature extraction function $[.]_P=<[.]_M,[.]_{tac}> : proof~patches \rightarrow M[\mathbb{Q}]_{5\times 6}$,
where $[.]_{tac}$ is an injective function that has been introduced to assign values to tactics.
We have defined two versions of  $[.]_{tac}$: one for Coq tactics and another for SSReflect tactics. 
In the SSReflect case, we divide the tactics into 7 groups and assign similar
values to each tactic in the group, see Table~\ref{tab:tactics}. 
Analogously for Coq tactics, cf. Appendix~\ref{sec:coqtactics}.


\begin{definition}[Proof-patch matrix]\label{def:ptm}
Given a Coq proof $C=((G_i,T_i))_{0\leq i\leq n}$, and a proof patch $p=((G_{i_0},T_{i_0}),\ldots,(G_{i_4},T_{i_4}))$ of $C$, 
the feature extraction function $[.]_P: proof~patches \rightarrow M[\mathbb{Q}]_{5\times 6}$ constructs the \emph{proof-patch matrix} $[p]_P$ 
as follows:

\begin{itemize}
 \item the $(j,0)$-th entry of $[p]_P$ is a 4-tuple $([T_{i_j}^1]_{tac}, [T_{i_j}^2]_{tac},[T_{i_j}^3]_{tac},[T_{i_j}^r]_{tac})$ 
 where $T_{i_j}^1, T_{i_j}^2$ and $T_{i_j}^3$ are the three first tactics of $T_{i_j}$, and $T_{i_j}^r$ is the list of the rest of tactics of $T_{i_j}$,
 \item the $(j,1)$-th entry of $[p]_P$ is the number of tactics appearing in $T_{i_j}$,
 \item the $(j,2)$-th entry of $[p]_P$ is a 4-tuple $([t_1]_{type},[t_2]_{type},[t_3]_{type},[t_{i_j}]_{type})$ where 
 $t_1, t_2$ and $t_3$ are the three first argument-types of $T_{i_j}$, and $t_{i_j}$ is the set of the rest of 
 the argument-types of $T_{i_j}$ (insensitive to order or repetition),
 \item the $(j,3)$-th entry of $[p]_P$ is a 4-tuple  $([l_{i_{j_1}}]_{term},[l_{i_{j_2}}]_{term},[l_{i_{j_3}}]_{term},[l_{i_j}]_{term})$
 where $l_{i_{j_1}}$, $l_{i_{j_2}}$ and $l_{i_{j_3}}$ are the three first lemmas applied in $T_{i_j}$ and $l_{i_j}$ is the list of the rest of lemmas 
used in $T_{i_j}$ (sensitive to order and repetition),
 \item the $(j,4)$-th entry of $[p]_P$ is a triple  $([s_1]_{term},[s_2]_{term},[s_3]_{term})$ where $s_1,s_2$ and $s_3$ are respectively the top, second, and third
 symbol of $G_{i_j}$,
 \item the $(j,5)$-th entry of $[p]_P$ is the number of subgoals after applying $T_{i_j}$ to $G_{i_j}$.
\end{itemize} 
\end{definition}

\begin{table}[t]
\centering
\begin{lstlisting}[frame=lines,mathescape,basicstyle=\tiny,breaklines=true]  
$\ast$ Bookkeeping ($b=\{$move:, move => $\}$): $[b_i]_{tac}=1+\sum_{j=1}^i \frac{1}{10\times 2^{j-1}}$ (where $b_i$ is the $i$th element of $b$).
$\ast$ Case and Induction ($c=\{$case, elim$\}$): $[c_i]_{tac}=2+\sum_{j=1}^i \frac{1}{10\times 2^{j-1}}$.
$\ast$ Discharge ($d=\{$apply, exact, congr$\}$): $[d_i]_{tac}=3+\sum_{j=1}^i \frac{1}{10\times 2^{j-1}}$.
$\ast$ Simplification ($s=\{$//, /=, //=$\}$): $[s_i]_{tac}=4+\sum_{j=1}^i \frac{1}{10\times 2^{j-1}}$.
$\ast$ Rewrite: $[$rewrite$]_{tac} = 5$. 
$\ast$ Forward Chaining  ($f=\{$have, suff, wlog$\}$): $[f_i]_{tac}=6+\sum_{j=1}^i \frac{1}{10\times 2^{j-1}}$.
$\ast$ Views and reflection  ($v=\{$move/, apply/, elim/, case/$\}$): $[v_i]_{tac}=7+\sum_{j=1}^i \frac{1}{10\times 2^{j-1}}$.
\end{lstlisting}
\caption{\scriptsize{\emph{Formulas computing the value of SSReflect tactics:} 
they serve to assign closer values to the tactics within each of the seven groups, and more distant numbers 
across the groups.  If a new tactic is defined, ML4PG automatically assigns a new number to it, using the next available natural number $n$ 
in the formula $n+\sum_{j=1}^i \frac{1}{10\times 2^{j-1}}$. 
}}\label{tab:tactics}
\end{table}

\begin{example}\label{example1}
Given the proof of Example~\ref{example0} and the proof-patch $((G_{i},T_{i}))_{0\leq i \leq 4}$, 
the top table of Table~\ref{tab:patches} shows its proof-patch matrix.
\end{example}

The \emph{proof-patch method} considers several proof-patches to collect information from a concrete proof. 
In particular, given a Coq proof $C=((G_i,T_i))_{0\leq i\leq n}$ , the proof $C$ can be split into patches $C_0,\ldots,C_m$ where 
$m=\lceil\frac{n}{5}\rceil+1$. The patches are defined as follows: $C_j=((G_j,T_j),\ldots,(G_{j+4},T_{j+4}))$ 
for $0\leq j < m$ (some patches can contain less than $5$ proof-steps); and $C_m=((G_{n-4},T_{n-4}),\ldots,(G_{n},T_{n}))$ -- the last patch captures 
the last five proof-steps.

\begin{example}
Using the proof-patch method, we  can split the proof presented in Example~\ref{example1} into three proof-patches $((G_i,T_i))_{0\leq i \leq 4}$, 
$((G_5,T_5),(G_6,T_6))$ and $((G_i,T_i))_{2\leq i \leq 6}$; the corresponding proof-patch matrices are given in Table~\ref{tab:patches}. 
\end{example}

The proof-patch method together with the feature function $[.]_P$ solve the two drawbacks of the old method~\cite{KHG13}: 
the new method captures information about the whole proof and the feature values are dynamically computed to assign close
values to similar terms, types, tactics and lemma statements used as tactic arguments.

We finish this section with a case study that illustrates 
the use of the proof-patch method and shows the differences with the results obtained with the old method~\cite{KHG13}. This case study concerns discovery of proof patterns in mathematical proofs
across formalisations of apparently disjoint mathematical theories: Linear Algebra, Combinatorics and Persistent Homology  (across 758 lemmas and 5~libraries).
In this scenario, we use statistically discovered proof patterns to advance the proof of a given ``problematic'' lemma. 
In this case,  a few initial steps in its proof are clustered against several mathematical libraries. 
We deliberately take lemmas belonging to very different SSReflect libraries. The lemma introduced in Example~\ref{example0} is a basic fact about
summations. Lemma~\ref{lem:nilpotent} states a result about \emph{nilpotent} matrices 
(a square matrix $M$ is \emph{nilpotent}
if there exists an $n$ such that $M^n=0$). Finally, Lemma~\ref{lem:fundamental} is a generalisation of the \emph{fundamental lemma of 
Persistent Homology}~\cite{HCMS12}.

\begin{lemma}\label{lem:nilpotent}
 Let $M$ be a square matrix and $n$ be a natural number such that $M^n=0$, then $(1-M)\times \sum\limits_{i=0}^{n-1} M^i = 1$.
\end{lemma}

\begin{lemma}\label{lem:fundamental}
Let $\beta_n^{k,l}:\mathbb{N} \times \mathbb{N} \times \mathbb{N} \rightarrow \mathbb{Z}$, then
$$\sum_{0\leq i \leq k} \sum_{l<j\leq m} (\beta_n^{i,j-1} - \beta_n^{i,j}) - (\beta_n^{i-1,j-1} - \beta_n^{i-1,j}) = \beta_n^{k,l} - \beta_n^{k,m}.$$
\end{lemma}

\begin{table}[t]
\centering
\tiny{
\begin{tabular}{|l||l|l|l|l|l|l|}
\hline
 & \emph{tactics} & \emph{n} & \emph{arg type} & \emph{arg} & \emph{symbols} & \emph{goals} \\
\hline
\hline
\emph{g1}& $([case]_{tac},0,0,0)$ & $1$  & $([nat]_{type},0,0,0)$  & $([Hyp]_{term},0,0,0)$ & $([\forall]_{term},[=]_{term},[sum]_{term})$ & $2$ \\
 \hline
  \emph{g2} & $([rewrite]_{tac},0,0,0)$ & $1$  & $([Prop]_{type},0,0,0)$  & $([big\_nat1]_{term},0,0,0)$ & $([=]_{term},[\sum]_{term},[-]_{term})$& $0$ \\
 \hline
  \emph{g3} & $([rewrite]_{tac},0,0,0)$ & $1$  & $([Prop]_{type},$ & $([surB]_{term},$  & $([=]_{term},[+]_{term},[-]_{term})$& $1$ \\
 & & &$[Prop]_{type},$ & $[big\_nat\_recr]_{term}$ & & \\
 & & &$[Prop]_{type},$ & $[big\_nat\_recl]_{term}$ & & \\
 & & &$[Prop]_{type})$ & $[EL]_{term})$ & & \\
 \hline
   \emph{g4} & $([move:]_{tac},0,0,0)$ & $1$  & $([Prop]_{type},0,0,0)$ &  $([eq\_refl]_{term},0,0,0)$  & $([=]_{term},[+]_{term},[-]_{term})$ & $1$ \\
 \hline 
  \emph{g5} & $([rewrite]_{tac},0,0,0)$ & $1$  & $([Prop]_{type},0,0,0)$  & $([subr\_eq0]_{term},0,0,0)$ & $([=]_{term},[+]_{term},[-]_{term})$ & $1$ \\
 \hline
  \end{tabular}
  
  \vspace{.2cm}
  
\begin{tabular}{|l||l|l|l|l|l|l|}
\hline
 & \emph{tactics} & \emph{n} & \emph{arg type} & \emph{arg} & \emph{symbols} &  \emph{goals} \\
\hline
\hline
\emph{g1} & $([move/]_{tac},[\texttt{->}]_{tac},0,0)$ & $2$  & $([Prop]_{type},0,0,0)$ &  $([eq\_refl]_{term},0,0,0)$  & $([=]_{term},[+]_{term},[-]_{term})$ & $1$ \\
 \hline
  \emph{g2} & $([rewrite]_{tac},0,0,0)$ & $1$  & $([Prop]_{type},0,0,0)$  & $([subr\_eq0]_{term},0,0,0)$ & $([=]_{term},[+]_{term},[-]_{term})$ & $1$ \\
 \hline
  \end{tabular}
  
  \vspace{.2cm}
  
\begin{tabular}{|l||l|l|l|l|l|l|}
\hline
 & \emph{tactics} & \emph{n} & \emph{arg type} & \emph{arg} & \emph{symbols} & \emph{goals} \\
\hline
\hline
\emph{g1}&  $([rewrite]_{tac},0,0,0)$ & $1$  & $([Prop]_{type},$ & $([surB]_{term},$  & $([=]_{term},[+]_{term},[-]_{term})$& $1$ \\
 & & & $[Prop]_{type},$ & $[big\_nat\_recr]_{term}$ & & \\
 & & & $[Prop]_{type},$& $[big\_nat\_recl]_{term}$ & & \\
 & & & $[Prop]_{type})$& $[EL]_{term})$ & & \\
 \hline
  \emph{g2} & $([move:]_{tac},0,0,0)$ & $1$  & $([Prop]_{type},0,0,0)$ &  $([eq\_refl]_{term},0,0,0)$  & $([=]_{term},[+]_{term},[-]_{term})$ & $1$ \\
 \hline
  \emph{g3} & $([rewrite]_{tac},0,0,0)$ & $1$  & $([Prop]_{type},0,0,0)$  & $([subr\_eq0]_{term},0,0,0)$ & $([=]_{term},[+]_{term},[-]_{term})$ & $1$ \\
 \hline
\emph{g4}& $([move/]_{tac},[\texttt{->}]_{tac},0,0)$ & $2$  & $([Prop]_{type},0,0,0)$  & $([eqP]_{term},0,0,0)$ & $([=]_{term},[+]_{term},[-]_{term})$ & $1$ \\
 \hline
  \emph{g5} & $([rewrite]_{tac},0,0,0)$ & $1$  & $([Prop]_{type},0,0,0)$  & $([sub0r]_{term},0,0,0)$ & $([=]_{term},[+]_{term},[-]_{term})$ & $0$ \\
 \hline
  \end{tabular}  
  }

 \caption{\scriptsize{\emph{Proof-patch matrices for the proof of Example~\ref{example1}.} \textbf{Top.} Proof-patch matrix of the patch $((G_i,T_i))_{0\leq i\leq 4}$. 
 \textbf{Centre.} Proof-patch matrix of the patch $((G_i,T_i))_{5\leq i\leq 6}$ (rows that are not included in the table are filled with zeroes).
 \textbf{Bottom.} Proof-patch matrix of the patch $((G_i,T_i))_{2\leq i\leq 6}$.  
Where we use notation $EL$, ML4PG gathers the lemma names: (\lstinline?addrC?, \lstinline?addrC?~, \lstinline?subr_sub?, \ldots).}}\label{tab:patches}
 \end{table}

When proving Lemma~\ref{lem:nilpotent}, the user may call ML4PG after completing a few standard proof steps: apply induction and solve the base case using rewriting.
At this point it is difficult, even for an expert user, 
to get the intuition that he can reuse the proofs from Example~\ref{example0} and Lemma~\ref{lem:fundamental}. There are several reasons for this.
First of all, the formal proofs of these lemmas are in different libraries; then, it is difficult to establish a conceptual connection among them. Moreover,
although the three lemmas involve summations,
the types of the terms of those summations are different. Therefore, search based on types or keywords would not help. Even search  
of all the lemmas involving summations does not provide a clear suggestion, since there are more than $250$ lemmas -- a considerable number for handling them manually.

%
%

However, if only the lemmas from Example~\ref{example0} and Lemma~\ref{lem:fundamental} are suggested when proving Lemma~\ref{lem:nilpotent}, the expert would be able to
spot the following common proof pattern.

 
\begin{PS}\label{ps:math}
 
\emph{
	Apply case on $n$.
  \begin{enumerate}
   \item Prove the base case (a simple task).
   \item Prove the case $0<n$:
     \begin{enumerate}
     \item expand the summation,
     \item cancel the terms pairwise,
     \item the terms remaining after the cancellation are the first and the last one. 
     \end{enumerate}
  \end{enumerate}
}

\end{PS}

%
%

Using the method presented in~\cite{KHG13}, if ML4PG was invoked during the proof of Lemma~\ref{lem:nilpotent}, it would suggest the lemmas from Example~\ref{example0} and 
Lemma~\ref{lem:fundamental}. However, 5 irrelevant lemmas 
 about summations would also be suggested (irrelevant in the sense that they do not follow Proof Strategy~\ref{ps:math}). The cluster containing just the two desired lemmas could be obtained after
increasing the \emph{granularity} value~\cite{KHG13} -- a statistical ML4PG parameter that can be adjusted by the user to obtain more precise clusters. The new version of ML4PG suggests four proof fragments, all following Strategy~\ref{ps:math}; without needing to adjust granularity.

The new method brings two improvements:
(1) the number of suggestions is increased and (2) 
the clusters are more accurate. The proof-patch method considers fragments of proofs that are deep in the proof and were not considered before; therefore, it can find lemmas (more precisely patches of lemmas) that were not
included previously in the clusters. In our case study, ML4PG suggests two additional interesting proof fragments. The first one is an intermediate patch of the proof of 
Lemma~\ref{lem:fundamental}; then, two patches are suggested from this 
lemma: the proof-patch of the inner sum, and the proof-patch of the outer sum (both of them following Proof Strategy~\ref{ps:math}). The following lemma is also 
suggested.

\begin{lemma}\label{lem:nilpotent2}
Let $M$ be a nilpotent matrix, then there exists a matrix $N$ such that $N \times (1-M)=1$. 
\end{lemma}

At the first sight, the proof of this lemma is an unlikely candidate to follow  Proof Strategy~\ref{ps:math}, since the statement of the lemma
does not involve summations. However, inspecting its proof, we can see that it uses $\sum_{i=0}^{n-1} M^i$ as witness for $N$ and 
then follows Proof Strategy~\ref{ps:math}. In this case, ML4PG suggests the patch from the last five proof-steps that correspond to 
the application of Proof Strategy~\ref{ps:math}.

The new numbering of features produces more accurate clusters removing the irrelevant lemmas. In particular, using the default settings, 
ML4PG only suggests the lemma from Example~\ref{example0}, the two patches from Lemma~\ref{lem:fundamental} and the last patch from 
Lemma~\ref{lem:nilpotent2}. If the granularity is increased, the last patch from 
Lemma~\ref{lem:nilpotent2} is the only suggestion -- note that this is the closest lemma to Lemma~\ref{lem:nilpotent}.

\section{Conceptualisation of the statistical results}\label{sec:conceptualisation}

In the previous section, we have seen how ML4PG can be used to find families of proofs following 
a common proof pattern. However, the output provided by clustering algorithms is just a set
of similar patches with no hints of why these proof-patches are deemed similar. 
In this section, we present our approach to facilitate the understanding of proof patterns.

The first problem that we address is the discovery of the key features that were taken
into account during the cluster formation. This is a well-known problem in machine-learning 
known as \emph{feature selection}~\cite{Weka}. From a given set of features, feature selection algorithms 
generate different subsets of features and create a ranking of feature subsets.
ML4PG uses the \emph{correlation-based feature subset selection} algorithm implemented in Weka~\cite{Weka}
-- the machine-learning toolbox employed by ML4PG to extract the relevant features.

\begin{figure}[t]
\centering 
\includegraphics[scale=.19]{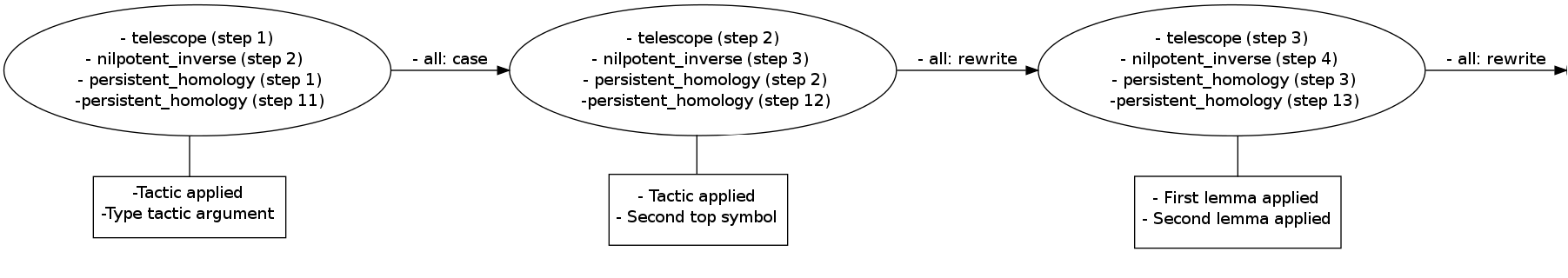}
\caption{\scriptsize{\emph{Fragment of the automaton 
 corresponding to the cluster of four lemma fragments described in the case study of Section~\ref{sec:recurrent}} (the whole automaton can be seen in Appendix~\ref{sec:automaton}). 
It shows correspondence between certain proof steps of lemmas 
\texttt{telescope} (for the lemma of Example~\ref{example0}), \texttt{nilpotent\_inverse} (for Lemma~\ref{lem:nilpotent2}), and \texttt{persistent\_homology} 
(for Lemma~\ref{lem:fundamental}). Square boxes denote feature correlation where it exists. 
To be compact, we  hide full lemma statements, tactic and auxiliary lemma names, symbols, etc., but they can be shown by ML4PG. 
}}\label{fig:automata}
\end{figure}

\begin{example}\label{ex:correlation-features}
In the case study presented at the end of the previous section, the relevant features were:

$-$ The tactics applied in the first and second steps of the proof-patches (\lstinline?case? and \lstinline?rewrite? respectively).

$-$ The type of the argument of the tactic applied in the first step of the proof-patches (\lstinline?nat?).

$-$ The second top symbol of the goal in the second step of the proof-patches (the $\sum$ symbol). 

$-$ The first and second auxiliary lemmas applied in the the third step of the proof-patches. The first auxiliary lemma (\lstinline?sumrB?) is used to expand the summations and is 
common to all the proofs. There are two different \lq\lq{}second\rq\rq{} auxiliary lemmas that occur in the proofs of Lemma~\ref{lem:fundamental} and~\ref{lem:nilpotent2} (\lstinline?big_nat_recr? and
\lstinline?big_nat_recl? -- both extract elements of a summation); however, the recurrent feature extraction process have assigned similar values to them.
The ``automaton'' of Example~\ref{ex:correlation-features} is depicted in Figure~\ref{fig:automata} and Appendix~\ref{sec:automaton}.
\end{example}

ML4PG uses this information and produces an automaton-shape representation for discovered 
proof-patterns and the correlated features that determined the patterns. 
Our graphical representation is simpler than other works where automata are used to represent models that are inferred from proof traces~\cite{GWR14}.  

Generally, given a cluster of proof patches ${\cal C}$, we have an automaton $A$ with $5$ consecutive states. The $i$th state of $A$ is labelled
with the list of $i$th goals in the proof-patches contained in ${\cal C}$. The transitions between the $i$th and $i+1$th states are given by 
the $i$th tactic of each proof-patch of ${\cal C}$. If two or more tactics belong to the same group (see Figure~\ref{tab:tactics});
they are merged in a unique transition; otherwise, the tactics will be shown as different transitions. In addition, each state is annotated with 
features 
whose correlation determined the cluster.

%


\section{Conclusions}\label{sec:conclusions}

We have presented three techniques to enhance the quality of ML4PG results. \emph{Term clustering} adds a new 
functionality to ML4PG: the user can receive suggestions about families of similar definitions, types and lemma shapes (in fact, any Coq terms). 
The \emph{proof-patch method} is employed to analyse the properties of the patches that constitute a proof. The whole syntax of Coq libraries is now subject to \emph{recurrent clustering}, which 
 increases the number and accuracy of families of similar proofs suggested by ML4PG. Finally,
the \emph{automaton-like representation} facilitates the interpretation of clusters of similar proof-patches. 

Further improvements in \emph{accuracy} (e.g. including proof contexts into the analysis) and \emph{conceptualisation} for clustered terms and proofs are planned. The families of similar proofs and terms can be the basis to
apply symbolic techniques to, for instance, infer models from proof traces~\cite{GWR14} or generate auxiliary results using
mutation of lemmas~\cite{lpar13}. The incorporation of these techniques will help in the goal pursued by ML4PG: make the 
proof development easier.



\bibliographystyle{plain}
\bibliography{itp14}

\begin{thebibliography}{10}

\bibitem{AspertiGCTZ04}
A.~Asperti et~al.
\newblock {A Content Based Mathematical Search Engine: Whelp}.
\newblock In {\em TYPES'04}, volume 3839 of {\em LNCS}, pages 17--32, 2006.

\bibitem{Bishop}
C.~Bishop.
\newblock {\em Pattern Recognition and Machine Learning}.
\newblock Springer, 2006.

\bibitem{Coq}
{\textsc{Coq} development team}.
\newblock {The \textsc{Coq} Proof Assistant Reference Manual, version 8.4pl3}.
\newblock Technical report, 2013.

\bibitem{CoquandH88}
T.~Coquand and G.~Huet.
\newblock The calculus of constructions.
\newblock {\em Information and Computation}, 76(2/3):95--120, 1988.

\bibitem{CoPa89}
T.~Coquand and C.~Paulin-Mohring.
\newblock Inductively defined types.
\newblock In {\em Colog'88}, volume 417 of {\em LNCS}, pages 50--66, 1990.

\bibitem{GarillotEtAl09}
F.~Garillot et~al.
\newblock Packaging mathematical structures.
\newblock In {\em TPHOLs'09}, volume 5674 of {\em LNCS}, 2009.

\bibitem{FCT}
G.~Gonthier.
\newblock Formal proof - the four-color theorem.
\newblock {\em Notices of the American Mathematical Society},
  55(11):1382--1393, 2008.

\bibitem{SSReflect}
G.~Gonthier and A.~Mahboubi.
\newblock {An introduction to small scale reflection}.
\newblock {\em Journal of Formalized Reasoning}, 3(2):95--152, 2010.

\bibitem{GWR14}
T.~Gransden et~al.
\newblock {Using Model Inference for Proof Development}.
\newblock Technical report, 2014.

\bibitem{Weka}
M.~Hall et~al.
\newblock {The WEKA Data Mining Software: An Update}.
\newblock {\em SIGKDD Explorations}, 11(1):10--18, 2009.

\bibitem{HCMS12}
J.~Heras et~al.
\newblock {Computing Persistent Homology within Coq/SSReflect}.
\newblock {\em ACM Transactions on Computational Logic}, 14(4), 2013.

\bibitem{lpar13}
J.~Heras et~al.
\newblock {Proof-Pattern Recognition and Lemma Discovery in ACL2}.
\newblock In {\em LPAR-19}, volume 8312 of {\em LNCS}, pages 389--406, 2013.

\bibitem{HK12}
J.~Heras and E.~Komendantskaya.
\newblock {ML4PG: downloadable programs, manual, examples}, 2012--2013.
\newblock \url{www.computing.dundee.ac.uk/staff/katya/ML4PG/}.

\bibitem{CICM13}
J.~Heras and E.~Komendantskaya.
\newblock {ML4PG in Computer Algebra Verification}.
\newblock In {\em CICM'13}, volume 7961 of {\em LNCS}, pages 354--358, 2013.

\bibitem{HK14}
J.~Heras and E.~Komendantskaya.
\newblock {Recycling Proof Patterns in Coq: Case Studies}.
\newblock {\em Journal Mathematics in Computer Science, accepted}, 2014.

\bibitem{lpar-urban}
C.~Kaliszyk and J.~Urban.
\newblock {Lemma Mining over HOL Light}.
\newblock In {\em LPAR-19}, volume 8312 of {\em LNCS}, pages 503--517, 2013.

\bibitem{KHG13}
E.~Komendantskaya et~al.
\newblock {Machine Learning for Proof General: interfacing interfaces}.
\newblock {\em Electronic Proceedings in Theoretical Computer Science},
  118:15--41, 2013.

\bibitem{mash}
D.~K\"uhlwein et~al.
\newblock {MaSh: Machine Learning for Sledgehammer}.
\newblock In {\em ITP'13}, volume 7998 of {\em LNCS}, pages 35--50, 2013.

\bibitem{UrbanSPV08}
J.~Urban et~al.
\newblock {MaLARea SG1- Machine Learner for Automated Reasoning with Semantic
  Guidance}.
\newblock In {\em IJCAR'08}, volume 5195 of {\em LNCS}, pages 441--456, 2008.

\end{thebibliography}

\pagebreak

\appendix

\section{Formula for Gallina tokens}\label{sec:gallinasyntax}

We split Gallina tokens into the following groups.

\begin{itemize}
 \item Group 1: \lstinline?forall?, \lstinline?->?.
 \item Group 2: \lstinline?fun?,
 \item Group 3: \lstinline?let?, \lstinline?let fix?, \lstinline?let cofix?.
 \item Group 4: \lstinline?fix?, \lstinline?cofix?.
 \item Group 5: \lstinline?@?,
 \item Group 6: \lstinline?match?, \lstinline?if?.
 \item Group 7: \lstinline?:=?, \lstinline?=>?, \lstinline?is?. 
 \item Group 8: \lstinline?Inductive?, \lstinline?CoInductive?.
 \item Group 9: \lstinline?exists?, \lstinline?exists2?.
 \item Group 10: \lstinline?:?, \lstinline?:>?, \lstinline?<:?, \lstinline?%?.

\end{itemize}

The formula for the $j$th Gallina token of the $n$th group is given by the formula $$- (n + \sum_{i=0}^j \frac{1}{10\times 2^{i-1}})$$

\newpage

\section{Groups of Coq tactics and number assignment}\label{sec:coqtactics}


Table~\ref{tab:coqtactics} splits the Coq tactics into different groups. The formula used in the function $[.]_{tactic}$ 
to compute the value of a Coq tactic is given by 
$i+\sum_{j=0}^k\frac{1}{10 \times 2^{j-1}}$ where $i$ is the group of the tactic and $k$ is the position 
of the tactic in that group (cf. right side of Table~\ref{tab:coqtactics}). 

\begin{table}[h]
\centering
\begin{tabular}{l||l}

Group & Tactics of the group\\

\hline
\hline
Group 1:           &  \lstinline?exact, eexact, assumption, eassumption, ?\\
Applying theorems &  \lstinline?refine, apply, eapply, simple apply, lapply, ?\\
\hline
Group 2:           &  \lstinline?constructor, split, exists, left, right, ?\\
Managing inductive constructors & \lstinline?econstructor, esplit, eexists, eleft, eright?\\
\hline
Group 3:           &  \lstinline?intro, intros, clear, rever, move, rename,  ?\\
Managing local context & \lstinline?set, remember, pose, decompose?\\
\hline
Group 4:            &  \lstinline?assert, cut, pose, specialize, generalize, ?\\
Controlling proof flow & \lstinline?evar, instantiate, admit, absurd, ?\\
& \lstinline?contradition, contradict, exfalso?\\
\hline
Group 5:            & \lstinline?destruct, case, ecase, simple destruct,?\\
Case analysis and induction & \lstinline?induction, einduction, elim, eelim, ?\\
 & \lstinline?simple induction, double induction, ?\\
     & \lstinline?dependent induction, functional induction,?\\
  & \lstinline? discriminate, injection, fix, cofix, ?\\
    & \lstinline? case_eq, elimtype?\\
\hline
Group 6:          &  \lstinline?rewrite, erewrite, cutrewrite, replace, ?\\
Rewriting expressions &  \lstinline?reflexivity, symmetry, transitivity, subst, ?\\
&  \lstinline?stepl, change?\\
\hline
Group 7:          &  \lstinline?cbv, compute, vm_compute, red, hnf, simpl, ?\\
Performing computations  &  \lstinline?unfold, fold, pattern, conv_tactic?\\
\hline
Group 8:           &  \lstinline?auto, trivial, eauto, autounfold, ?\\
Automation       &  \lstinline?autorewrite?\\
\hline
Group 9:           &  \lstinline?tauto, intuition, rtauto, firstorder, ?\\
Decision procedures & \lstinline?congruence?\\
\hline
Group 10 &  \lstinline?decide equality, compare, simplify_eq,?\\
Equality & \lstinline?esimplify_eq?\\
\hline
Group 11 &  \lstinline?inversion, dependent inversion,?\\
Inversion & \lstinline?functional inversion, quote?\\
\hline
Group 12 & \lstinline?classical_left, classical_right?\\
Classical tactics & \lstinline??\\
\hline
Group 13 & \lstinline?omega, ring, field, fourier?\\
Automatizing & \\
\hline
Group 14 & Rest of Coq tactics\\
\hline
\end{tabular}
\caption{{\scriptsize \emph{Groups of Coq tactics}}}\label{tab:coqtactics}
\end{table}

\newpage

\section{Automaton}\label{sec:automaton}

\begin{figure}[h]
    \centering
\includegraphics[scale=0.28]{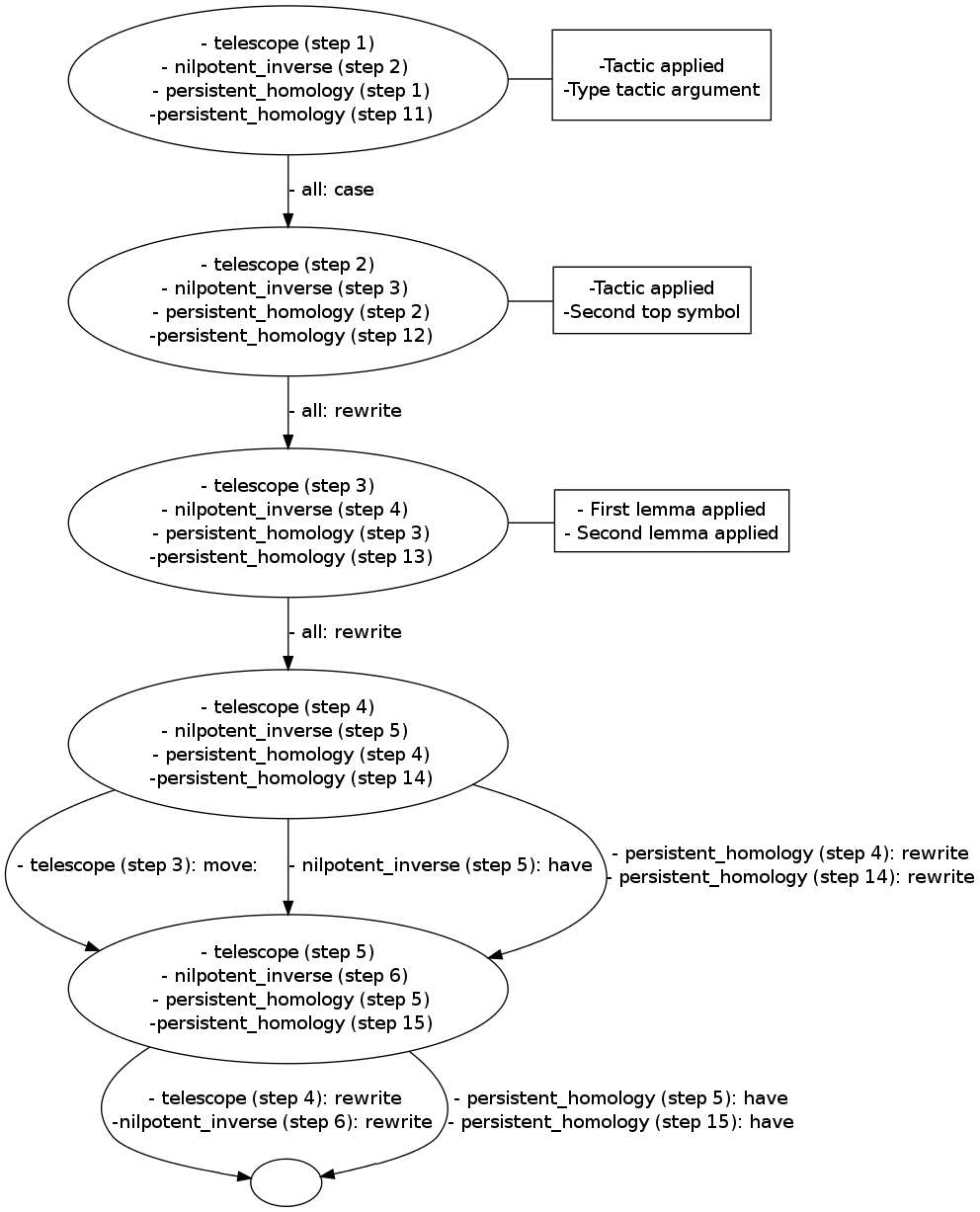} 
\caption{\scriptsize{\emph{Automaton corresponding to the proof cluster of four lemma fragments described in the case study of Section~\ref{sec:recurrent}.} The automaton shows the five proof steps, of which the first three are shown to influence the cluster formation.
It uses Lemma names: \texttt{telescope} for the lemma of Example~\ref{example0}, \texttt{nilpotent\_inverse} for Lemma~\ref{lem:nilpotent2}, and \texttt{persistent\_homology} 
for Lemma~\ref{lem:fundamental}.
  In addition to Lemma names, ML4PG can show lemma statements, and it can provide details of the \lq\lq{}Tactic applied\rq\rq{}, \lq\lq{}Type tactic argument\rq\rq{}, \lq\lq{}Second top symbol\rq\rq{}, \lq\lq{}First/Second lemma applied\rq\rq{} fields.}}
\end{figure}

\end{document}